\documentclass[a4paper,11pt]{article}
\usepackage{pos}
\usepackage{multirow}
\usepackage{float}
\usepackage{amsfonts}
\usepackage{dsfont}

\title{Heavy hadron spectrum from 2+1+1 flavor MILC lattices}

\author*[a]{Sabiar Shaikh}
\author[b]{Protick Mohanta}
\author[b,c]{M. Padmanath}
\author[a,c]{Subhasish Basak}

\affiliation[a]{School of Physical Sciences, National Institute of Science Education and Research, An OCC of Homi Bhabha National Institute, Jatni 752050, India}

\affiliation[b]{The Institute of Mathematical Sciences, Chennai 600113, India}
\affiliation[c]{Homi Bhabha National Institute, Training School Complex,
Anushakti Nagar, Mumbai  400094, India}

\emailAdd{sabiars@niser.ac.in}
\emailAdd{protickm@imsc.res.in}
\emailAdd{padmanath@imsc.res.in}
\emailAdd{sbasak@niser.ac.in}

\abstract{ We study the mass spectra and various mass differences of heavy hadrons containing one or more bottom quarks using MILC's $N_f = 2+1+1$ HISQ gauge ensembles at three lattice spacings. For the valence quarks, we employ a combination of lattice actions: the NRQCD action is used for bottom quarks, the anisotropic Clover action for charm quarks, and the $O(a)$-improved Wilson--Clover action for strange and lighter (up/down) quarks. Heavy hadron operators with at least one bottom quark are constructed by considering all possible combinations with charm, strange, and light quarks corresponding to various quantum numbers.}

\FullConference{The 42nd International Symposium on Lattice Field Theory (LATTICE 2025)\\
2-8 November 2025\\
Tata Institute of Fundamental Research, Mumbai, India\\}

\begin{document}
\maketitle

\section{Introduction}
Lattice QCD has been extensively applied to $B$ physics, especially for determining decay constants, mixing parameters, and meson mass differences relevant to CKM matrix elements~\cite{Davies:2011tfc}. Lattice QCD provides a first-principles approach to investigate the masses, splittings, and other properties of these baryons. Numerous studies have examined heavy baryons with one to three bottom quarks using various light quark actions~\cite{Mathur:2018epb, Meinel:2010pw, Brown:2014ena}. The spectroscopy of heavy hadrons containing one or more bottom quarks provides a crucial window into the non-perturbative dynamics of QCD and tests lattice formulations for heavy quarks. Previous lattice studies have successfully used nonrelativistic QCD (NRQCD) to describe bottom quarks~\cite{Lepage:1992tx}, yet achieving higher precision and controlling systematic uncertainties remain challenges, especially in light of ongoing experimental progress in heavy-flavor physics. In this work, we build on earlier investigations by simulating bottom quarks with NRQCD, charm quarks with an anisotropic clover action, and strange/light quarks with the Wilson--Clover action, enabling a systematic study of bottom-hadron spectroscopy with controlled discretization effects.\\
We present lattice QCD results for the ground-state spectra of hadrons containing at least one bottom quark, computed on MILC $N_f = 2+1+1$ HISQ ensembles at two lattice spacings, with a finer ensemble in progress. Interpolating operators are constructed for bottom mesons and baryons with all combinations of bottom, charm, strange, and light quarks. For baryons, we focus on positive-parity states with spins $J = \tfrac{1}{2}$ and $J = \tfrac{3}{2}$, and extract ground-state masses and mass differences. This study extends previous lattice work and represents a step toward precision determinations of bottom-hadron spectra from first principles, providing valuable input for current and future heavy-flavor experiments.

\section{Quark action and tuning}
\subsection{NRQCD action for the bottom quark}

In lattice QCD calculations of hadrons containing bottom quarks on relatively coarse lattices, NRQCD  formulation is the most widely used approach~\cite{Lepage:1992tx}. The bottom quark inside a hadron is highly nonrelativistic, with $v^2 \sim 0.1$, as indicated by the comparison of bottomonium masses to the quark mass. For bottom hadrons with lighter valence quarks, the bottom quark velocity is even smaller, justifying the use of a nonrelativistic effective field theory. NRQCD thus provides a suitable framework for studying bottom quarks and is expected to remain the preferred method until finer lattices with $a m_b < 1$ become widely available.\\
In NRQCD formulation, the upper and lower components of the Dirac spinor decouple and the bottom quark is described by two-component spinor fields. The NRQCD Lagrangian up to  order $\mathcal O(v^4)$ is~\cite{Bodwin:1994jh}
 \begin{equation}
 \mathcal{L} =~ \psi^\dagger(D_4 + H_0 + \delta H)\psi ~+~ \xi^\dagger
 (D_4 - \overline{H_0} - \overline{\delta H})\xi ,
  \label{eq1}
\end{equation}
where $\psi$ and $\xi$ are the bottom quark and antiquark fields, respectively. The Hamiltonian $H_0$ contains the leading $\mathcal{O}(v^2)$ term and $\delta H=\sum \delta H_i$ contains the $\mathcal{O}(v^4)$ and $\mathcal{O}(v^6)$ terms with coefficients $c_i$. Similarly for $\overline H_0$ and $\overline{\delta H}$.

The various terms in the quark and antiquark Hamiltonians are~\cite{HPQCD:2011qwj, Bodwin:1994jh}
\begin{eqnarray}   \nonumber
 H_0 &=& - \frac{\Delta^2}{2m_b} ~=~ \overline{H_0} \\   \nonumber
 \delta H_1 &=& -c_1 \frac{(\Delta^2)^2}{8m_b^3} ~=~ \overline{\delta H_1} \\ \nonumber
 \delta H_2 &=& c_2 \frac{i}{8m_b^2}{\bf(\nabla\cdot\tilde{E}
 -\tilde{E}\cdot\nabla)} ~=~ -\overline{\delta H_2} \\   \nonumber
 \delta H_3 &=& -c_3 \frac{1}{8m_b^2}{\bf\sigma.(\tilde{\nabla}\times
 \tilde{E} -\tilde{E}\times\tilde{\nabla})} ~=~ -\overline{\delta H_3}\\   \nonumber
 \delta H_4 &=& -c_4 \frac{1}{2m_b}{\bf \sigma\cdot \tilde{B}} ~=~
 \overline{\delta H_4} \\   \nonumber
 \delta H_5 &=& c_5 \frac{\Delta^4}{24m_b} ~=~ \overline{\delta H_5}\\  
 \delta H_6 &=& -c_6\frac{(\Delta^2)^2}{16nm_b^2} ~=~ \overline{\delta H_6}.
 \label{eq2}
\end{eqnarray}
Here $n$ is the factor in $\delta H_6$ to ensure numerical stability at small $am_b$ with $n > 3/2m_b$~\cite{Mohanta:2019mxo}. On the lattice, the heavy quark Green's function evolution is approximated by~\cite{Lepage:1992tx} 
\begin{equation}
G_\psi(x,t+1;0,0)=\Big(1-{\delta H\over 2}\Big)\Big(1-{H_0\over {2n}}
\Big)^nU^{\dagger}_t(x,t)
\Big(1-{H_0\over 2n}\Big)^n \Big(1-{\delta H\over 2}\Big) 
G_\psi(x,t;0,0),
\label{eq3}
\end{equation}
with the initial condition 
\begin{equation}
G_\psi(x,t;x_0,t_0) = \delta_{x,x_0} \, \delta_{t,t_0} \, \mathbb{I}.
\label{eq4}
\end{equation}
 The heavy-quark propagator $G_\psi$
propagates only forward in time. Similarly, the heavy antiquark evolution equation with $G_\xi$ evolving backwards in time is given by
\begin{equation}
G_\xi(x,t-1;0,0)=\Big(1-{\overline{\delta H}\over 2}\Big)
\Big(1-{\overline{H_0}\over2n}\Big)^nU_t(x,t-1)
\Big(1-{\overline{H_0}\over 2n}\Big)^n
\Big(1-{\overline{\delta H}\over 2}\Big)G_\xi(x,t;0,0),
 \label{eq5}
\end{equation}
using the initial condition 
\begin{equation}
G_\xi(x,t;x_0,t_0) = -\delta_{x,x_0} \, \delta_{t,t_0} \, \mathbb{I}.
\label{eq6}
\end{equation}

The open bottom system, $B_s$ meson, is used for tuning the bottom quark. From the energy-momentum relation~\cite{HPQCD:2011qwj}
\begin{equation}
E({\bf P}) = M_1 + \frac{{\bf P}^2}{2M_2},
 \label{eq7}
\end{equation}
where \(M_1\) is the pole mass, and the kinetic mass \(M_2\) of $B_s$ is tuned to match the PDG value of 5366.93 MeV. The hyperfine splitting, defined as
\begin{equation}
\Delta M_{B_s} = M_{B_s^*}^{\rm rest} - M_{B_s}^{\rm rest} = 48.5~\text{MeV},
 \label{eq8}
\end{equation}
provides a consistency check when using the pole masses of $B_s$ and $B_s^*$ at zero momentum. The NRQCD coefficients \(c_1, c_2, c_3, c_5, c_6\), the exponent \(n\), and the tadpole factor \(u_0\) are taken from HPQCD results~\cite{HPQCD:2011qwj}. The tuning parameters are the bare quark mass $m_b$ and the coefficient $c_4$. $c_4$ is the coefficient of order $\mathcal O ({1 / m_b})$ term which gives the leading contribution in $\delta H$ and control the hyperfine splitting.

\subsection{Anisotropic clover action for charm quark}
In this work, we use the anisotropic clover-improved Wilson action or Relativistic Heavy Quark (RHQ) action for the charm quark. The anisotropic Clover action modifies the standard Wilson-Clover formulation by introducing distinct spatial ($a_s$) and temporal ($a_t$) lattice spacings, leading to an anisotropy factor $\nu = a_s / a_t$. This setup enhances temporal resolution, which is crucial for heavy quark systems, while maintaining manageable lattice sizes. The RHQ action $S_{\textmd{RHQ}}$ is given by~\cite{Christ:2006us, RBC:2012pds}
\begin{eqnarray}
{S_{\textmd{RHQ}} \over m_0+1+3\nu} &=& \sum_n \bar{\psi}(n)\psi(n) -
\kappa\Big[\sum_{n}\bar{\psi}(n)(1-\gamma_0)U_0(n)\psi(n+\hat{0})\nonumber\\
&+& \sum_{n}\bar{\psi}(n)(1+\gamma_0)U^\dagger_0(n-\hat{0})
\psi(n-\hat{0}) + \nu\sum_{n,i}\bar{\psi}(n)(1-\gamma_i)U_i(n)
\psi(n+\hat{i})\nonumber\\
&+& \nu\sum_{n,i}\bar{\psi}(n)(1+\gamma_i)U^\dagger_i(n-\hat{i})
\psi(n-\hat{i})\Big] -\kappa c_P\sum_{n,\mu<\nu}
\bar{\psi}(n)\sigma_{\mu\nu}F_{\mu\nu}(n)\psi(n),
 \label{eq9}
\end{eqnarray}
where  $\kappa =  \dfrac {1}{ 2(m_0 +1+3\nu)}$, $c_P$ is the clover coefficient, $\sigma_{\mu\nu} = \frac{i}{2} [\gamma_\mu, \gamma_\nu]$ is the antisymmetric tensor, and $F_{\mu\nu}(n)$ represents the discretized gluon field strength and taken to be hermitian. The determination of the RHQ action parameters $\{ m_0, c_P, \nu \}$ is performed using the $D_s$ meson correlations, as this choice mitigates both discretization effects and uncertainties from chiral extrapolation. The value of $m_0$ is tuned by matching lattice-calculated meson masses to experimental values. Specifically, the spin-averaged mass of the $D_s$ and $D_s^*$ mesons is used
\begin{equation}
\bar {M}_{D_s} = \frac{1}{4} M_{D_s} + \frac{3}{4} M_{D_s^*}.
 \label{eq10}
\end{equation}
The PDG value of $\bar {M}_{D_s}$ is approximately 2.076 GeV.
The clover coefficient $c_P$ is adjusted so that the hyperfine splitting
\begin{equation}
\Delta M_{D_s} = M_{D_s^*} - M_{D_s},
 \label{eq11}
\end{equation}
matches the PDG value of 143.8 MeV.
The anisotropy coefficient $\nu$ is adjusted so that the meson dispersion relation satisfies the relativistic condition
\begin{equation}
E^2(\vec{p}) = M^2 + \vec{p}^{\,2},
 \label{eq12}
\end{equation}
ensuring that the speed of light is correctly normalized on the lattice to $c^2=1$.
We used $|\vec{p}|={2\pi \over L}$ in tuning the RHQ action parameters.

\subsection{Clover action for strange, up/down quark}

To simulate strange and light (up/down) quarks, we employ the $\mathcal{O}(a)$-improved Wilson--Clover fermion action, which improves upon the original Wilson action by including a clover term that systematically removes leading discretization errors at $\mathcal{O}(a)$. The fermion action is expressed as
\begin{eqnarray}
S_{\textmd{clover}} &=& \sum_n \bar{\psi}(n)\psi(n) - \kappa_s\Big[
\sum_{n,\mu}\bar{\psi}(n)(1-\gamma_\mu)U_\mu(n)\psi(n+\hat{\mu})\nonumber\\
&+& \sum_{n,\mu}\bar{\psi}(n)(1+\gamma_\mu)U^\dagger_\mu(n-\hat{\mu})
\psi(n-\hat{\mu})\Big] -\kappa_s c_{\textmd{SW}}\sum_{n,\mu<\nu}
\bar{\psi}(n)\sigma_{\mu\nu}F_{\mu\nu}(n)\psi(n) 
 \label{eq13}
\end{eqnarray}
where $\kappa_s =\dfrac {1} {2(m + 4)}$ and $c_{\text{SW}}$ is the clover coefficient. In order to determine the $\kappa_s$, we have used the fictitious $\eta_s$ meson.  From chiral perturbation
theory \cite{Mohanta:2019mxo}, its mass is estimated to be ~$m_{\eta_s} = $ 688.5 MeV~\cite{Dowdall:2013rya}. The bare quark mass is controlled via the hopping parameter $\kappa_s$, which is related to the bare mass through
\begin{equation}
m a = \frac{1}{2\kappa_s} - 4.
 \label{eq14}
\end{equation}
 For light quarks (up/down), the hopping parameter $\kappa_{u/d}$ is tuned to obtain a lower pion mass $m_\pi$. The meson masses are extracted from two-point correlation functions computed on the lattice. Similarly, by tuning the values of $\kappa_s$, we set the $\eta_s$ meson mass to be 688.5 MeV. The clover coefficient $c_{\text{SW}}$ is tuned using the tadpole improvement prescription~\cite{FermilabLattice:2011njy}
\begin{equation}
 c_{\text{SW}} = \frac{1}{u_0^3},
  \label{eq15}
\end{equation}
where $u_0$ is the tadpole improvement factor, typically defined via the fourth root of the average plaquette.
This methodology provides a controlled and efficient framework for simulating strange and light quarks in lattice QCD using the improved Clover action.

\section{Bottom baryon operators}
To construct the baryon interpolating operators, we consider
different quark compositions depending on the number of heavy quarks. 
For triple, double and single bottom baryons, the operators are defined in Table~\ref{bottom_baryon_ops}.

\begin{table}[h]
\centering
\begin{tabular}{|c|c|}
\hline
\textbf{Baryon Type} & \textbf{Interpolating Operator} \\
\hline
Triple-bottom & $\left( \mathcal{O}^{hhh}_k \right)_\alpha = \epsilon_{abc} ({Q^a}^T C \gamma_k Q^b) Q^c_\alpha$ \\[1mm]
\hline
\multirow{3}{*}{Double-bottom} & $(\mathcal{O}^{hhl}_k)_\alpha = \epsilon_{abc} ({Q^a}^T C \gamma_k Q^b) l^c_\alpha$ \\ 
& $(\mathcal{O}^{hlh}_k)_\alpha = \epsilon_{abc} ({Q^a}^T C \gamma_k l^b) Q^c_\alpha$ \\
& $(\mathcal{O}^{hlh}_5)_\alpha = \epsilon_{abc} ({Q^a}^T C \gamma_5 l^b) Q^c_\alpha$ \\
\hline
\multirow{3}{*}{Single-bottom} & $(\mathcal{O}^{hl_1l_2}_5)_\alpha = \epsilon_{abc} ({l_1^a}^T C \gamma_5 l_2^b) Q^c_\alpha$ \\
& $(\mathcal{O}^{hl_2l_1}_k)_\alpha = \epsilon_{abc} ({Q^a}^T C \gamma_k l_2^b) l^c_{1\alpha}$ \\
& $(\mathcal{O}^{hl_2l_1}_5)_\alpha = \epsilon_{abc} ({Q^a}^T C \gamma_5 l_2^b) l^c_{1\alpha}$ \\
\hline
\end{tabular}
\caption{Interpolating operators for bottom baryons.}
\label{bottom_baryon_ops}
\end{table}
$Q$ denotes NRQCD bottom quarks~\cite{Leskovec:2019ioa}, $l$ is light or charm quarks, $C$ is the charge-conjugation matrix, $\alpha$ is the spinor index, and 
\begin{equation}
Q=\Theta (t - t')\begin{pmatrix} \psi_h \\ 0 \ \end{pmatrix} - \begin{pmatrix} 0 \\ \chi_h \ \end{pmatrix}  \Theta (-t+t').
 \label{eq16}
\end{equation}
Here, in the first term $t'$ is positive, whereas in the
second term $t'$ is negative. These operators are constructed to respect
color antisymmetry and the required spin structure, allowing for the
extraction of ground-state baryon masses with different heavy-quark
content.

\section{Simulation Details}
The simulations have been performed using three ensembles of MILC $N_f=2+1+1$ HISQ lattices~\cite{HPQCD:2011qwj}, the details of which are given in Table 2. The tuned parameters reported for $32^3 \times 96$ lattices are only for 50 configurations. The parameters are tuned for a smaller subset of configurations, and the production runs are ongoing. Here we present a status update of this ongoing effort. All measurements have been carried out using the point sources. 
\begin{table}[h]
\centering
\begin{tabular}{|c|c|c|c|c|c|c|}
\hline
$\beta = {10} /{g^2}$ & $am_l$ & $am_s$ & $am_c$ &
$L^3 \times T$ & $a$ (fm) & $N_{cfg}$ \\
\hline
5.80 & 0.013 & 0.065 & 0.838 & $16^3 \times 48$ & 0.15 & 700 \\
\hline
6.00 & 0.0102 & 0.0509 & 0.635 & $24^3 \times 64$ & 0.12 & 700 \\
\hline
6.30 & 0.0074 & 0.037 & 0.440 & $32^3 \times 96$ & 0.09 & 50 \\
\hline
\end{tabular}
\caption{MILC configurations used in this work. The gauge
coupling is $\beta$, lattice spacing a. $am_l$, $am_s$, and $am_c$ are the light (up and down taken to have the same mass), strange and charm sea quark masses in lattice units, respectively, and the lattice size is $L^3 \times T$. The $N_{cfg}$ is the
number of configurations used in this work.}
\label{tab:ensembles}
\end{table}

\subsection{Strange quark tuning}
For the strange-quark, the hopping parameter
$\kappa_s$ has been tuned separately for each lattice to get the ChiPT $\eta_s$ mass as discussed above. The clover
coefficient has been set to $c_{\mathrm{SW}} = u_0^{-3}$ following Ref.~\cite{FermilabLattice:2011njy}, with the tadpole improvement factor
$u_0$ taken from Ref.~\cite{MILC:2010pul}. The resulting parameters are
given in  Table~\ref{tab:strange} for different lattices.

\begin{table}[h]
\centering
\begin{tabular}{|c|c|c|}
\hline
$L^3 \times T$ & $\kappa_s$ & $c_{\mathrm{SW}}$ \\
\hline
$16^3 \times 48$ & 0.14101  & 1.598 \\
\hline
$24^3 \times 64$ & 0.140015 & 1.552 \\
\hline
$32^3 \times 96$ & 0.138613 & 1.497 \\
\hline
\end{tabular}
\caption{Strange-quark hopping parameters and clover coefficients used in the simulations.}
\label{tab:strange}
\end{table}

\subsection{Charm-quark tuning}
The charm-quark mass has been tuned for the RHQ
action Eq.~\ref{eq9} by matching the spin-averaged $D_s$ mass Eq.~\ref{eq10}  and the hyperfine
splitting Eq.~\ref{eq11} with the PDG values. The charm-quark tuning parameters are given in Table~\ref{tab:Ds_tuning}.
\begin{table}[h]
\centering
\begin{tabular}{|c|c|c|c|c|c|c|}
\hline
$L^3 \times T$ & $m_0$ & $c_P$ & $\nu$ & $\bar M_{D_s}$ (MeV) & $\Delta M_{D_s}$ (MeV) & $c^2$ \\
\hline
$16^3 \times 48$ & 0.6647 & 2.5822 & 1.2488 & 2076 (9) & 144 (9) & 1.0003 \\
\hline
$24^3 \times 64$ & 0.2033 & 1.9264 & 1.2168 & 2076 (7) & 144 (7) & 0.9999 \\
\hline
$32^3 \times 96$ & $-0.556$ & 0.36 & 1.52 & 2077 (15) &144 (15) & 1.0028  \\
\hline
PDG &  &  &  & 2076.24 & 143.8 &  \\
\hline
\end{tabular}
\caption{Charm-quark tuning using the $\bar M_{D_s}$ and $\Delta M_{D_s}$.}
\label{tab:Ds_tuning}
\end{table}

\subsection{Bottom-quark tuning}
 The bottom-quark mass has been tuned for the NRQCD
action Eq.~\ref{eq1} by matching the kinetic mass of the $B_s$  Eq.~\ref{eq7} and the
hyperfine splitting $\Delta M_{B_s}$ Eq.~\ref{eq8} with the PDG values. The bottom-quark tuning parameters are given in Table~\ref{tab:Bs_tuning}. 
\begin{table}[h]
\centering
\begin{tabular}{|c|c|c|c|c|}
\hline
$L^3 \times T$ & $m_b$ & $c_4$ & $B_s$ kinetic mass (MeV) & $\Delta M_{B_s}$ (MeV) \\
\hline
$16^3 \times 48$ & 3.8167 & 1.4456 & 5367 & 48.5 \\
\hline
$24^3 \times 64$ & 2.2498 & 1.0336 & 5367 & 48.5 \\
\hline
$32^3 \times 96$ & 1.44 & 0.9 & 5366  & 49.1  \\
\hline
PDG & & & 5366.93 & 48.5 \\
\hline
\end{tabular}
\caption{Bottom-quark tuning using the $B_s$ kinetic mass and $\Delta M_{B_s}$. The error in determining kinetic mass is around 2.5\%.}
\label{tab:Bs_tuning}
\end{table}

In the NRQCD formulation, absolute hadron masses are not directly
accessible due to the omission of the heavy quark mass term from
the action. The physical $B_s$ meson mass is reconstructed
by adding an ensemble-dependent energy offset to the pole mass
extracted from the two-point correlation functions. The kinetic masses reproduce
the experimental $B_s$ mass, while the offsets account for the missing
rest-mass contribution in NRQCD.  

\section{Results and Discussions}
\begin{figure}[H]
\centering
    \begin{minipage}[b]{0.48\textwidth}
    \includegraphics[width=\textwidth]{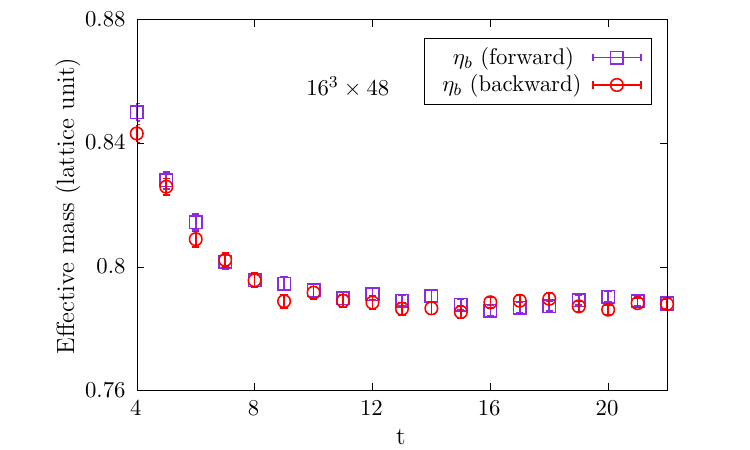}    
   \end{minipage}
\hspace{0.1cm}
  \begin{minipage}[b]{0.48\textwidth}
    \includegraphics[width=\textwidth]{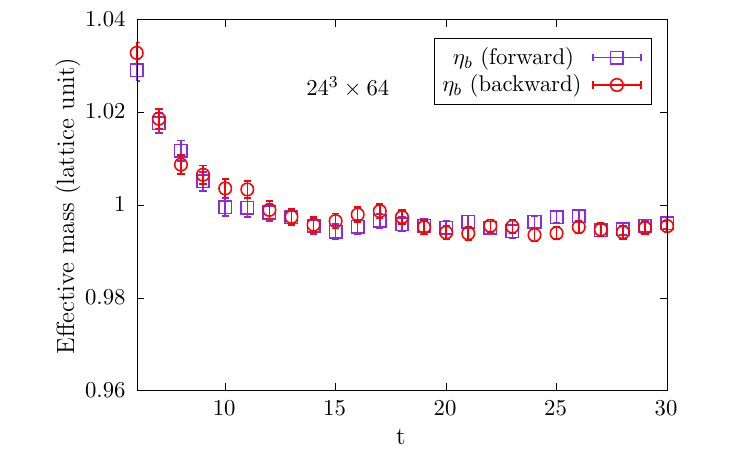}
    \end{minipage}
     \caption{Effective masses of forward and backward propagation of $\eta_b$, on $16^{3}\times48$ and $24^{3}\times64$ lattices, obtained for test of unitarity.}
      \label{commass16} 
 \end{figure}

Unitarity of the four-component NRQCD formulation was investigated by comparing effective masses extracted from correlation functions of forward and backward propagation of $\eta_b$. 
Within statistical uncertainties, the effective masses obtained from the forward and backward propagations are found
to agree over the plateau region, resulting in a consistent ground-state
extraction, indicating unitarity is preserved in the four-component NRQCD approach. The resulting plots are shown in Fig.~\ref{commass16} for $\eta_b$.

\begin{figure}[H]
\centering
  \begin{minipage}[b]{0.48\textwidth}
    \includegraphics[width=\textwidth]{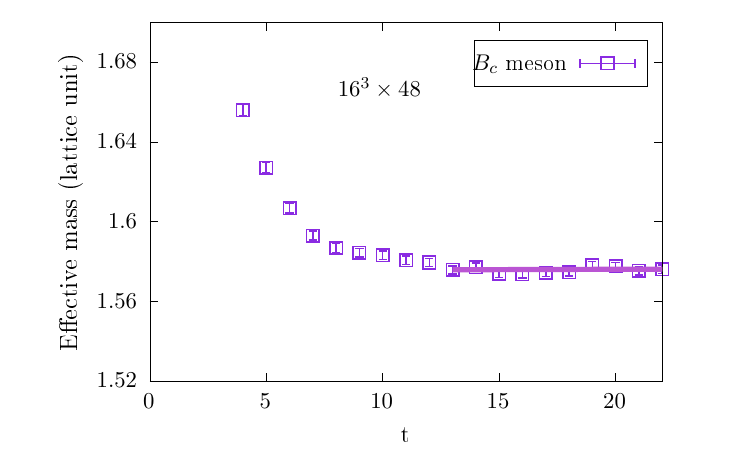}  
   \end{minipage}
\hspace{0.1cm}
  \begin{minipage}[b]{0.48\textwidth}
    \includegraphics[width=\textwidth]{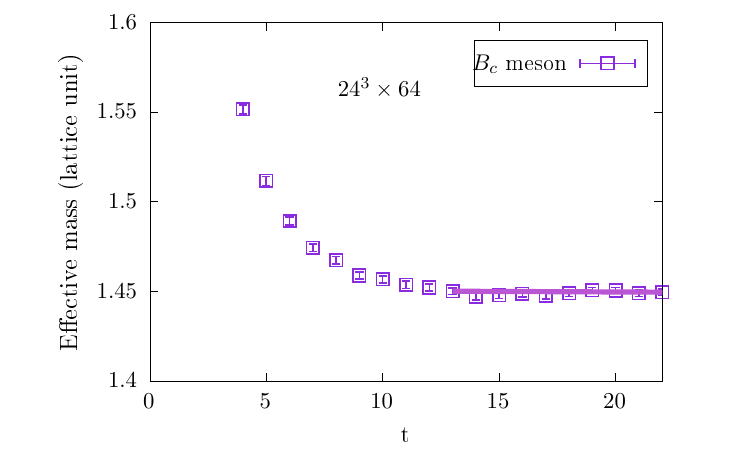}    
    \end{minipage}
     \caption{ Effective mass plots for $B_c$. The bands are placed over the fit range.}
      \label{effmass3}
 \end{figure}

 The $B_c$ meson, which contains both a bottom and a charm quark,
provides a stringent test of the simultaneous tuning of the NRQCD
action for the bottom quark and the RHQ
action for the charm quark. We compute the $B_c$ two-point correlation
functions on the $16^{3}\times48$ and $24^{3}\times64$ ensembles and extract the corresponding effective masses. The resulting effective
mass plots exhibit clear and stable plateaus over an extended range of
Euclidean time. This enables us to reliably extract the ground-state. A representative effective-mass results for the
two ensembles are shown in Fig.~\ref{effmass3}.

\begin{figure}[H]
\centering
  \begin{minipage}[b]{0.48\textwidth}
    \includegraphics[width=\textwidth]{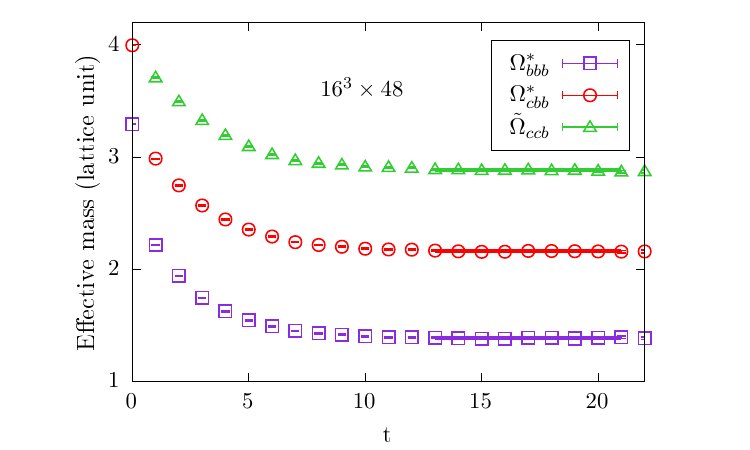}
   \end{minipage}
\hspace{0.1cm}
  \begin{minipage}[b]{0.48\textwidth}
    \includegraphics[width=\textwidth]{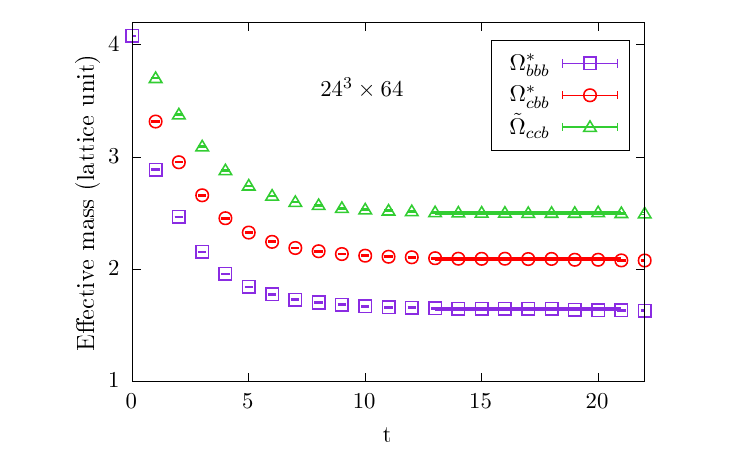}
    \end{minipage}
    \caption{$\Omega^*_{bbb}$, $\Omega^*_{cbb}$, and $\tilde\Omega_{ccb}$ effective masses for $16^3 \times 48$ and $24^3 \times 64$ ensembles.}  
    \label{effmass5}
 \end{figure}

We extract baryon masses from two-point correlation functions on the
$16^3 \times 48$ and $24^3 \times 64$ ensembles. The corresponding
effective masses exhibit reasonably long plateaus over
Euclidean time. A representative effective-mass plots for the two ensembles
are shown in Fig.~\ref{effmass5}, demonstrating stable and
consistent behavior across the plateau region.

\begin{figure}[H]
\centering
    \begin{minipage}[b]{0.48\textwidth}
    \includegraphics[width=\textwidth]{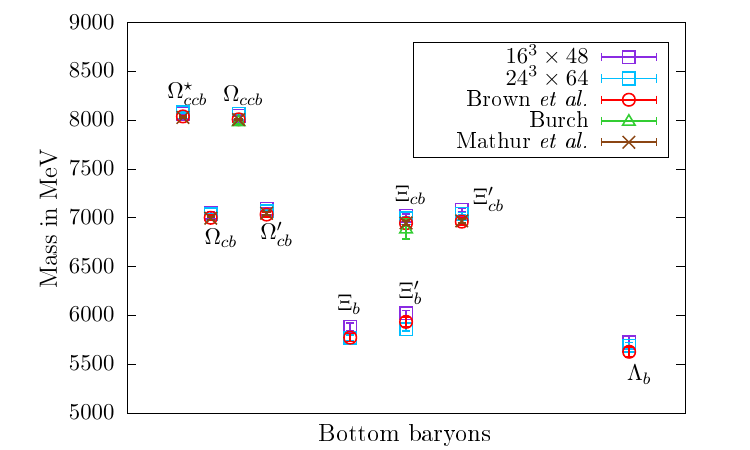}
   \end{minipage}
\hspace{0.1cm}
  \begin{minipage}[b]{0.48\textwidth}
    \includegraphics[width=\textwidth]{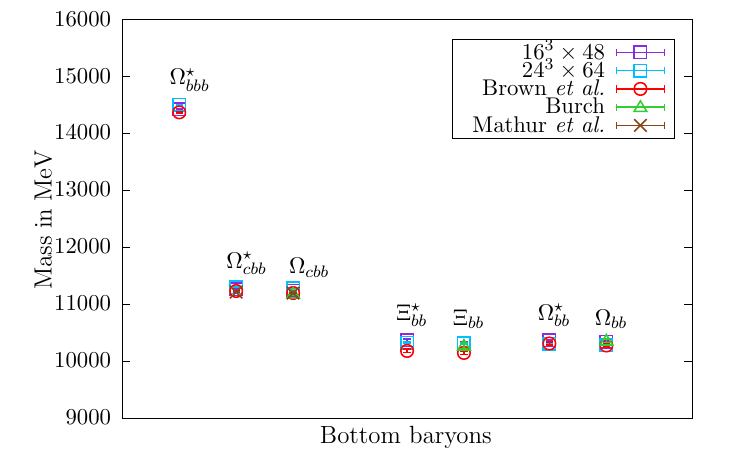}
    \end{minipage}
     \caption{Comparison of our single, double and triple bottom baryon spectra with Brown et al.~\cite{Brown:2014ena}, Burch~\cite{Burch:2015pka}, and Mathur et al.~\cite{Mathur:2018epb, Mathur:2002ce}.}  
     \label{spectra1}
 \end{figure}

The resulting baryon mass spectra was computed on the $16^3 \times 48$
and $24^3 \times 64$ ensembles. As shown in Fig.~\ref{spectra1},
our results for both single and multiple-bottom baryons are in
good agreement with previous lattice calculations, demonstrating
the consistency and reliability of our heavy-quark tuning and
methodology. The spectra exhibits the expected mass hierarchies and
splittings, providing a comprehensive benchmark for future
comparisons with experimental and lattice data.

\section{ Summary and Outlook}

In this talk, we give a status update of the lattice QCD study of heavy hadron mass spectra
containing one or more bottom quarks using $2+1+1$ flavor HISQ
ensembles generated by the MILC Collaboration.  Our calculations have been performed at two lattice spacings, with a third finer lattice spacing currently in progress, to get continuum extrapolation and improved control over discretization effects. A mixed-action approach is employed, using NRQCD for bottom quarks, RHQ action for charm quarks, and the
clover-improved Wilson action for strange and light quarks. The quark
masses are tuned carefully using known hadron masses, including
$\eta_s$, $D_s$, and $B_s$, to ensure accurate reproduction of
physical observables.

Our analysis focuses primarily on the ground-state masses of
single and multiple-bottom hadrons. The resulting spectra exhibit
stable effective-mass plateaus and show good agreement with
available experimental values and previous lattice calculations. This consistency confirms the reliability of our mixed-action setup for
heavy-quark physics. Future work will extend these studies to hyperfine splittings, mass differences, and additional heavy-baryon states. 
These investigations will lead to a more complete understanding of the heavy-hadron spectrum. 
They will also provide important benchmarks for both experimental measurements and theoretical models.

\section*{Acknowledgement}

We acknowledge financial support from the Department of Atomic Energy
(DAE), Government of India, and NISER, Bhubaneswar. We thank the MILC
Collaboration for providing the HISQ gauge ensembles used in this
work. Simulations have been carried out on the Bihan cluster of the School of Physical Sciences, NISER and the Kamet cluster at IMSc. 
MP gratefully acknowledges support from the Department
of Science and Technology, India, SERB Start-up Research Grant No. SRG/2023/001235.
We also thank Srijit Paul for helping with the HISQ gauge configurations.


\begin{thebibliography}{99}

\bibitem{Davies:2011tfc}
C.~Davies,
PoS \textbf{LATTICE2011}, 019 (2011)
doi:10.22323/1.139.0019
[arXiv:1203.3862 [hep-lat]].


\bibitem{Mathur:2018epb}
N.~Mathur, M.~Padmanath and S.~Mondal,
Phys. Rev. Lett. \textbf{121}, no.20, 202002 (2018)
doi:10.1103/PhysRevLett.121.202002
[arXiv:1806.04151 [hep-lat]].


\bibitem{Meinel:2010pw}
S.~Meinel,
Phys. Rev. D \textbf{82}, 114514 (2010)
doi:10.1103/PhysRevD.82.114514
[arXiv:1008.3154 [hep-lat]].

\bibitem{Brown:2014ena}
Z.~S.~Brown, W.~Detmold, S.~Meinel and K.~Orginos,
Phys. Rev. D \textbf{90}, no.9, 094507 (2014)
doi:10.1103/PhysRevD.90.094507
[arXiv:1409.0497 [hep-lat]].


\bibitem{Lepage:1992tx}
G.~P.~Lepage, L.~Magnea, C.~Nakhleh, U.~Magnea and K.~Hornbostel,
Phys. Rev. D \textbf{46}, 4052-4067 (1992)
doi:10.1103/PhysRevD.46.4052
[arXiv:hep-lat/9205007 [hep-lat]].

\bibitem{Bodwin:1994jh}
G.~T.~Bodwin, E.~Braaten and G.~P.~Lepage,
Phys. Rev. D \textbf{51}, 1125-1171 (1995)
[erratum: Phys. Rev. D \textbf{55}, 5853 (1997)]
doi:10.1103/PhysRevD.55.5853
[arXiv:hep-ph/9407339 [hep-ph]].


\bibitem{HPQCD:2011qwj}
R.~J.~Dowdall \textit{et al.} [HPQCD],
Phys. Rev. D \textbf{85}, 054509 (2012)
doi:10.1103/PhysRevD.85.054509
[arXiv:1110.6887 [hep-lat]].


\bibitem{Mohanta:2019mxo}
P.~Mohanta and S.~Basak,
Phys. Rev. D \textbf{101}, no.9, 094503 (2020)
doi:10.1103/PhysRevD.101.094503
[arXiv:1911.03741 [hep-lat]].


\bibitem{Christ:2006us}
N.~H.~Christ, M.~Li and H.~W.~Lin,
Phys. Rev. D \textbf{76}, 074505 (2007)
doi:10.1103/PhysRevD.76.074505
[arXiv:hep-lat/0608006 [hep-lat]].


\bibitem{RBC:2012pds}
Y.~Aoki \textit{et al.} [RBC and UKQCD],
Phys. Rev. D \textbf{86}, 116003 (2012)
doi:10.1103/PhysRevD.86.116003
[arXiv:1206.2554 [hep-lat]].


\bibitem{Dowdall:2013rya}
R.~J.~Dowdall, C.~T.~H.~Davies, G.~P.~Lepage and C.~McNeile,
Phys. Rev. D \textbf{88}, 074504 (2013)
doi:10.1103/PhysRevD.88.074504
[arXiv:1303.1670 [hep-lat]].


\bibitem{FermilabLattice:2011njy}
A.~Bazavov \textit{et al.} [Fermilab Lattice and MILC],
Phys. Rev. D \textbf{85}, 114506 (2012)
doi:10.1103/PhysRevD.85.114506
[arXiv:1112.3051 [hep-lat]].

\bibitem{Leskovec:2019ioa}
L.~Leskovec, S.~Meinel, M.~Pflaumer and M.~Wagner,
Phys. Rev. D \textbf{100}, no.1, 014503 (2019)
doi:10.1103/PhysRevD.100.014503
[arXiv:1904.04197 [hep-lat]].

\bibitem{MILC:2010pul}
A.~Bazavov \textit{et al.} [MILC],
Phys. Rev. D \textbf{82}, 074501 (2010)
doi:10.1103/PhysRevD.82.074501
[arXiv:1004.0342 [hep-lat]].


\bibitem{Burch:2015pka}
T.~Burch,
[arXiv:1502.00675 [hep-lat]].


\bibitem{Mathur:2002ce}
N.~Mathur, R.~Lewis and R.~M.~Woloshyn,
Phys. Rev. D \textbf{66}, 014502 (2002)
doi:10.1103/PhysRevD.66.014502
[arXiv:hep-ph/0203253 [hep-ph]].






\end{thebibliography}
\end{document}